\documentclass[12pt]{iopart}

\usepackage{iopams}

\usepackage{graphicx}
\usepackage{dcolumn}
\usepackage{bm}
\usepackage{times}

\newcommand{\Tc}{$T_c$}

\begin{document}

\title{Superconductivity in single crystals of LaFePO}

\author{J.\ J.\ Hamlin, R.\ E.\ Baumbach, D.\ A.\ Zocco, T.\ A.\ Sayles and M.\ B.\ Maple}

\address{Department of Physics and Institute for Pure and
Applied Physical Sciences, University of California, San Diego, La
Jolla, CA 92093}

\ead{mbmaple@ucsd.edu}

\begin{abstract}
Single crystals of the compound LaFePO were prepared using a flux
growth technique at high temperatures. Electrical resistivity
measurements reveal metallic behavior and a resistive transition
to the superconducting state at a critical temperature
$T_c\sim6.6$ K. Magnetization measurements also show the onset of
superconductivity near 6 K. In contrast, specific heat
measurements manifest no discontinuity at \Tc.  These results lend
support to the conclusion that the superconductivity is associated
with oxygen vacancies that alter the carrier concentration in a
small fraction of the sample, although superconductivity
characterized by an unusually small gap value can not be
ruled-out. Under applied magnetic fields, \Tc\ is suppressed
anisotropically for fields perpendicular and parallel to the
$ab$-plane, suggesting that the crystalline anisotropy strongly
influences the superconducting state.  Preliminary high-pressure
measurements show that \Tc\ passes through a maximum of nearly 14
K at $\sim 110$ kbar, demonstrating that significantly higher \Tc\
values may be achieved in the phosphorus-based oxypnictides.

\end{abstract}

\maketitle

\section{Introduction}
There has been a flurry of research activity following the recent
reports of superconductivity with high critical temperatures \Tc\
in the system LnFeAs[O$_{1-x}$F$_{x}$] where Ln is a lanthanide
\cite{kamihara_2008_1,takahashi_2008_1,fang_2008_2,chen_2008_1,ren_2008_1,sales_2008_1}.
To date, values of \Tc\ as high as 55 K have been reported for Ln
= Sm \cite{ren_2008_1}.  These compounds belong to a general class
of compounds with a layered crystal structure of the form LnFePnO
that were first synthesized by Jeitschko and coworkers with Pn = P
\cite{zimmer_1995_1} and As \cite{quebe_2000_1}. Superconductivity
in this series of materials was discovered in LaFePO in 2006 by
Kamihara \textit{et al.\ } \cite{kamihara_2006_1}, for which values
of \Tc\ that range from 3 K \cite{kamihara_2006_1} to 7 K
\cite{tegel_2008_1} have been reported. However, in a recent study
of polycrystalline materials, it was concluded that stoichiometric LaFePO
is metallic but non-superconducting \cite{mcqueen_2008_1} at
temperatures as low as 0.35 K.  In this paper, we report the
synthesis of single crystals of LaFePO. These crystals exhibit
superconducting transitions at 6.6 K and 6.0 K, according to
electrical resistivity and magnetic susceptibility measurements,
respectively. However, there is no specific heat jump at \Tc,
suggesting that only a small fraction of the sample is
superconducting.  The superconductivity appears to be a property
of the single crystals, since the resistively measured upper
critical field is quite anisotropic. It is possible that the
superconductivity is associated with oxygen vacancies that dope a
small fraction of the compound with charge carriers.  We present
measurements of the electrical resistivity, magnetic
susceptibility, and specific heat in the normal state. Preliminary
measurements of the pressure dependence of \Tc\ are also reported.

\section{Experimental Details}
\begin{figure}
    \begin{center}
        \includegraphics[width=12cm]{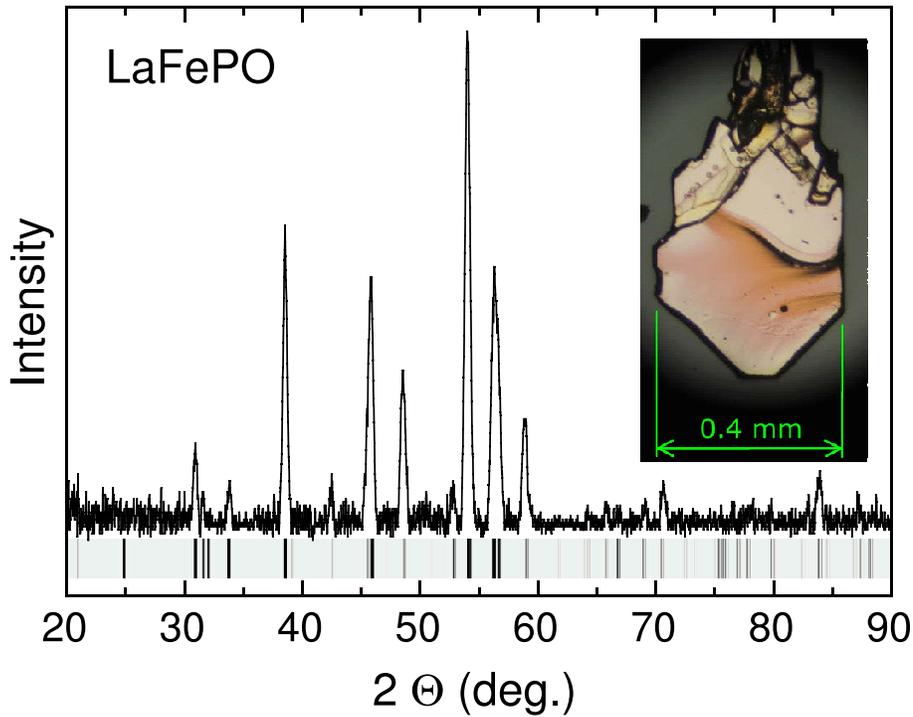}
    \end{center}
    \caption{Powder x-ray diffraction pattern for LaFePO. The $y$-axis is linear.  The vertical lines below the diffraction pattern indicate
    calculated Bragg peak positions. Inset: (color) Photograph shows a typical single crystal of LaFePO.}
    \label{fig:fig_xrd}
\end{figure}

Single crystals of LaFePO were grown from elements and elemental
oxides with purities $>99.9$\% in a molten Sn:P flux. The growths
took place over a 1 week period in quartz ampoules which were
sealed with 75 torr Ar at room temperature. The inner surface of
each quartz ampoule was coated with carbon by a typical pyrolysis
method. The starting materials were La, Fe$_2$O$_3$, P, and Sn,
which were combined in the molar ratios 9:3:6:80.5, similar to a
previous synthesis reported by Krellner and Geibel
\cite{geibel_2008_1} for CeRuPO single crystals. The Fe$_2$O$_3$
powder was dried for $\sim 10$ hours at 300 $^{\circ}$C before
weighing. The ampoule was heated to 1135 $^{\circ}$C at a rate of
35 $^{\circ}$C/hr, kept at this temperature for 96 hours, and then
rapidly cooled to 700 $^{\circ}$C. After removing the majority of
the flux by spinning the ampoules in a centrifuge, LaFePO single
crystal platelets of an isometric form with typical dimensions of
$\sim0.5 \times 0.5 \times 0.05$ mm$^3$ were collected and cleaned
in hydrochloric acid to remove the flux from the surface of the
crystals prior to measurements. After cleaning, the crystals were
observed to have a distinct gold color.  The platelets cleaved
easily in the $ab$-plane and were notably malleable, in contrast
to the cuprate superconductors.  A typical platelet is shown in
the inset of Fig.\ \ref{fig:fig_xrd}.

X-ray powder diffraction measurements were made using a Bruker D8
diffractometer with a non-monochromated Cu K$\alpha$ source to
check the purity and crystal structure of the LaFePO single
crystals. Due to their malleability, the crystals were difficult
to grind into a fine powder. Thus, the powder diffraction pattern
was generated from a collection of several crystals which
were cut into small pieces using a razor blade and then ground
into a coarse powder using a mortar and pestle.

\begin{figure}
    \begin{center}
        \includegraphics[width=12cm]{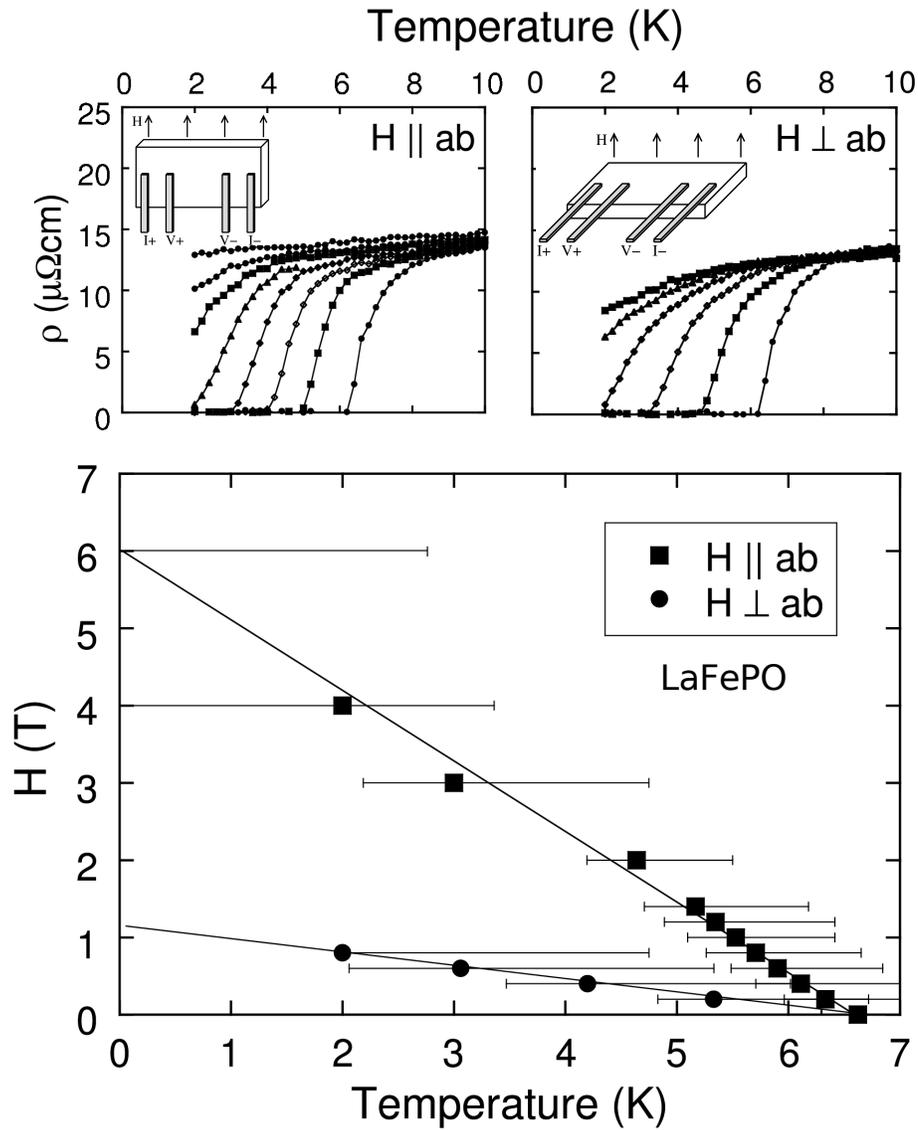}
    \end{center}
    \caption{Top panels: Electrical resistivity $\rho$ versus temperature
    $T$ data for LaFePO single crystal platelets for temperatures 2-10 K and magnetic
    fields $\|$ to the $ab$-plane: 0 T, 1.0 T, 2.0 T, 3.0 T, 4.0 T,
    5.0 T, 6.0 T, and 8.0 T (top left panel) and $\bot$ to the $ab$-plane: 0 T, 0.2 T, 0.4 T, 0.6 T,
    0.8 T, and 1.0 T (top right panel). Bottom panel: The upper critical field line $H_{c2}$ versus $T$ for
    $H$ $\|$ and $\bot$ to the $ab$-plane.}
    \label{fig:fig_RH}
\end{figure}

Electrical resistivity $\rho(T)$ measurements were performed in a
four-wire configuration with the current in the $ab$-plane, at
temperatures $T$ $=$ 2-300 K and magnetic fields $H$ $=$ 0-8 T
using a conventional $^4$He cryostat and a Quantum Design Physical
Properties Measurement System (PPMS). In order to explore the
anisotropy of the superconducting properties, the resistivity was
measured for 2-10 K with $H$ both parallel and perpendicular to
the crystal $ab$-planes, $\rho(T,H_{\|})$ and $\rho(T,H_{\bot})$,
respectively. For these measurements, the geometry was arranged
such that the current flowed in the $ab$-plane and was
perpendicular to $H$, as illustrated in the upper panels of Fig.\
\ref{fig:fig_RH} .

\begin{figure}
    \begin{center}
        \includegraphics[width=12cm]{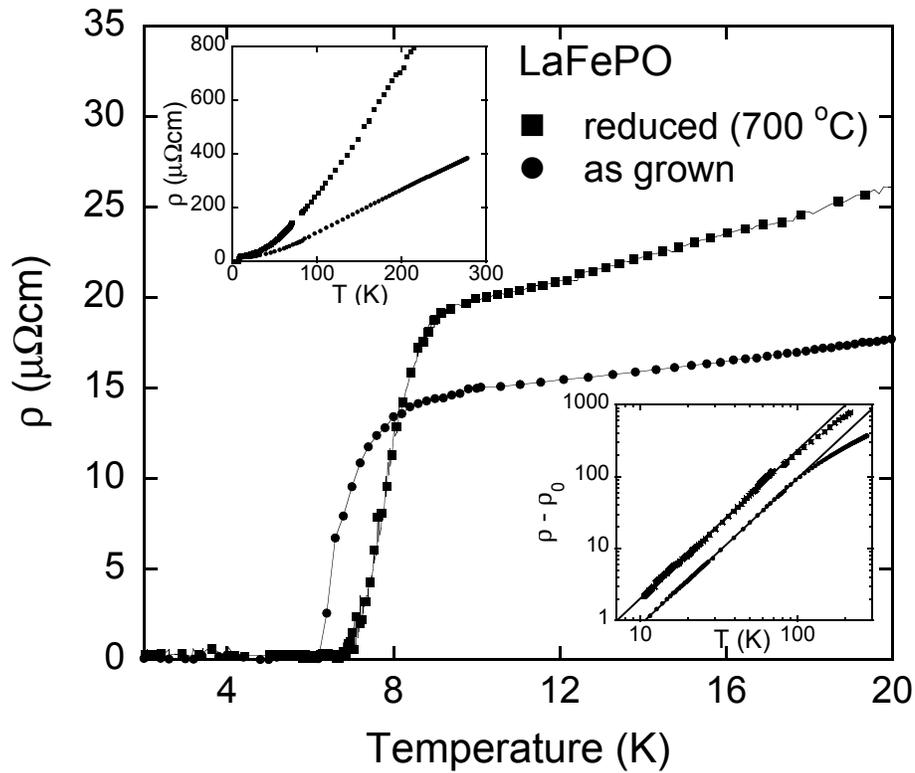}
    \end{center}
    \caption{Electrical resistivity $\rho$ versus temperature $T$ measured in the $ab$-plane for an
    as-grown single crystal and a crystal that was reduced in flowing argon at 700 $^0$C for $\sim 12$ hours.
    Left inset: Resistivity over the entire measured temperature
range showing $T^2$ behavior for $\sim$10-100 K and linear $T$ behavior
for $\sim$100-300 K. Right inset: Log-log plot of $\rho - \rho _0$
versus $T$.  The solid lines are fits to the data which demonstrate the $T^2$ behavior for 10-100 K for the as grown sample and 10-80 K for the reduced sample. }
    \label{fig:fig_R}
\end{figure}

DC magnetization $M(T,H)$ measurements were made using a Quantum
Design Magnetic Properties Measurement System (MPMS) in order to
probe both the superconducting and normal state properties of the
single crystal platelets. The specimens were mounted in
cotton-packed gelatin capsules with the $ab$-plane perpendicular
to the magnetic field. Multiple single crystal platelets were each
individually measured for 2-10 K and $H=5$ Oe under both zero
field cooled (ZFC) and field cooled (FC) conditions in order to
characterize batch homogeneity via variation in \Tc, which
was found to be minimal. Magnetization versus temperature at
$H=1.0$ T and 7-300 K was measured for a collection of several
hundred single crystal specimens ($m=13.06$ mg) in order to
characterize the normal state.

Specific heat $C(T)$ measurements were made for 2-300 K in a
Quantum Design PPMS semiadiabatic calorimeter using a heat-pulse
technique on the same collection of single crystals used for the
normal state magnetization measurements. The specimens were
attached to a sapphire platform with a small amount of Apiezon N
grease.

\begin{figure}
    \begin{center}
        \includegraphics[width=12cm]{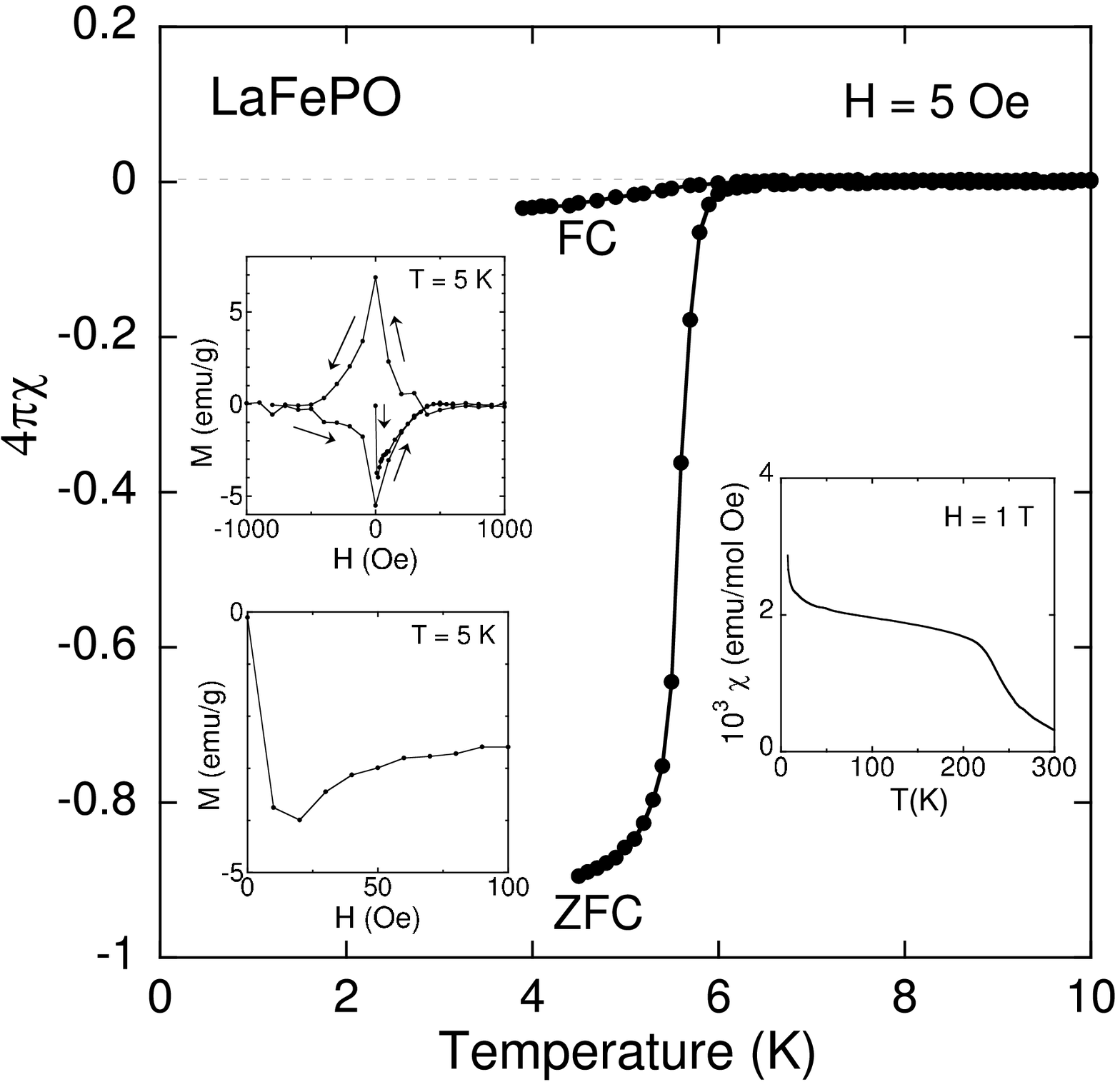}
    \end{center}
    \caption{DC magnetic susceptibility $M/H$ versus temperature $T$ taken in $H=5$ Oe for a single crystal
    specimen under zero-field-cooled (ZFC) and field-cooled (FC) conditions. Left insets: $M$ versus $H$ exhibiting both
    flux expulsion (Meissner) and flux penetration (vortex) superconducting states as expected for a type-II superconductor.
Right inset: $M/H$ versus temperature $T$ for 7-300 K and $H=1.0$
T.}
    \label{fig:fig_SQUID}
\end{figure}

A single high-pressure experiment was performed using a
mechanically loaded commercial diamond anvil cell (DAC),
manufactured by Kyowa Seisakusho Ltd.  The diamond anvils were
beveled from 500 to 250 $\mu$m tips.  One of the diamonds contains
six deposited tungsten microprobes encapsulated in high-quality
homepitaxial diamond. The fabrication of ``designer'' diamonds is
described in Ref. \cite{weir_2000_1}. The gasket was made from a
200 $\mu$m thick MP35N foil pre-indented to 40-50 $\mu$m with a
130 $\mu$m diameter hole drilled through the center of the
pre-indented region using an electrical discharge machine (EDM).
Two 5 $\mu$m diameter ruby spheres were loaded into the hole in
the gasket and the remaining space in the hole was filled with
$\sim 5$ small pieces of LaFePO crystal.  Because the sample is in
direct contact with the metallic gasket, the measured resistance
results from a combination of the sample and gasket resistivity.  Due to the geometry of the leads, the measured resistance generally does not drop completely to zero when the sample becomes superconducting.
However, such a configuration is sufficient for locating the \Tc\
of the sample. Pressure was adjusted and determined at room
temperature, using the fluorescence spectrum of the ruby spheres
and the calibration of Chijioke \textit{et al.\ }
\cite{chijioke_2005_1}. A delrin spacer is used to minimize
changes in pressure during cooling to low $T$.  Additional details
of the DAC technique are described in Ref. \cite{jeffries_2006_1}.

\section{Results and Discussion}
Shown in Fig.\ \ref{fig:fig_xrd} is the x-ray diffraction pattern
for a collection of LaFePO single crystals. The diffraction
pattern conforms to the characteristic tetragonal phase of LaFePO
consisting of P-Fe$_2$-P layers of edge sharing Fe-P octahedra
separated by La-O$_2$-La sheets in which the Fe atoms form
two-dimensional square nets. The diffraction pattern appears to be
in good agreement with previous reports
\cite{zimmer_1995_1,kamihara_2006_1,mcqueen_2008_1} and exhibits
no impurity contributions down to the 10\% level. Measurements on
a mosaic of several single crystals reveal that the platelets grow
with their large faces parallel to the 00$l$ crystal planes.

The electrical resistivity data, shown in Fig.\ \ref{fig:fig_R},
reveal metallic behavior where $\rho(T)$ decreases with decreasing
$T$ until it drops abruptly to zero near the superconducting
transition temperature  $T_c \sim 6.6$ K. This transition
temperature is defined as the temperature where $\rho(T)$ drops to
50\% of its extrapolated normal state value. The transition width
$\Delta T_c=1.3$ K is taken as the difference in the temperatures
where $\rho(T)$ drops to 10\% and 90\% of the extrapolated normal
state value. For $\sim$100-300 K, $\rho$($T$) has an approximately
linear $T$ dependence which evolves into a quadratic form for
$\sim$10-100 K, as shown in the left inset to Fig.\ \ref{fig:fig_R}. Fits
over this temperature range show that the data are well described by
the expression $\rho(T) = \rho_0 + AT^2$ where $\rho_0 \sim 14$
$\mu \Omega$cm and $A = 9.46 \times 10^{-3} \mu \Omega$cm/K$^2$.
The residual resistivity ratio $RRR=\rho(300 K)/\rho(0)=32$
reflects the high quality of the LaFePO single crystal. Also shown
in Fig.\ \ref{fig:fig_R} are results for a platelet which was
reduced at 700 $^{\circ}$C in flowing Ar for 24 hours. For this
specimen, $RRR=58$, $T_c = 7.8$ K, and the $\Delta T_c=1.6$ K,
indicating that the superconducting state may be enhanced by
reduction of oxygen concentration.  Fits to the data for the
reduced sample give $\rho_0 \sim 16.7$ $\mu \Omega$cm and $A =
2.44 \times 10^{-2} \mu \Omega$cm/K$^2$.

Shown in Fig.\ \ref{fig:fig_RH} are $\rho(T,H_{\|})$ and $\rho(T,H_{\bot})$ data for 2-10 K and 0-8 T which reveal pronounced anisotropy in the upper critical field curve $H_{c2}(T)$. Again, \Tc\ is defined as 50\% of the extrapolated normal state value while the transition width is taken as the difference between the temperatures where $\rho(T)$ drops to 10\% and 90\% of the extrapolated normal state value. At zero $T$, the anisotropy is quantified by the ratio $H_{c2}^{\|}(T)/H_{c2}^{\bot}(T)\sim 5.2$. This result suggests that the layered structure strongly influences the in-plane and inter-plane transport behavior.  The Clogston-Chandrasekhar \cite{clogston_1962_1,chandrasekhar_1962_1} Pauli-paramagnetic limiting field at zero temperature is given in units of tesla by $H_{p0}\equiv1.84 T_c$, which for $T_c=6.6$ K gives $H_{p0}\sim 12.1$ T, a value well above the extrapolated zero temperature critical field line illustrated in Fig.\ \ref{fig:fig_RH}.  Thus, the upper critical field is limited by orbital depairing for both $H \| ab$ and $H \bot ab$.  The Ginzburg - Landau coherence lengths parallel and perpendicular to the $ab$-plane, $\xi _{\|}$ and $\xi _{\bot}$, respectively, can be estimated from the slopes of $H_{c2}^{\|}$ and $H_{c2}^{\bot}$ near \Tc\ \cite{ginsberg_1989_1}, \textit{i.\ e.}, 

\begin{equation}
(dH_{c2}^{\bot}/dT) _{T_c} = -\Phi_0 / 2\pi T_c \xi _{\|} ^{2}
\end{equation}

and

\begin{equation}
(dH_{c2}^{\|}/dT) _{T_c} = -\Phi_0 / 2\pi T_c \xi _{\bot} \xi _{\|},
\end{equation}
where $\Phi_0=hc/2e=2.07\times 10^{-7}$ G$\cdot$cm$^2$ is the flux quantum.  From the values $(dH_{c2}^{\bot}/dT)_{T_c} = -1700$ Oe/K and $(dH_{c2}^{\|}/dT)_{T_c} = -8600$ Oe/K we obtain $\xi _{\|}= 170$ \AA{} and $\xi _{\bot}= 34$ \AA{}.

In an effort to determine whether the superconductivity is a bulk
phenomenon, zero-field-cooled (ZFC) and field-cooled (FC)
measurements of the magnetic susceptibility were made in a field
of 5 Oe. A plot of the ZFC and FC magnetic susceptibility through
the superconducting transition is shown in Fig.\
\ref{fig:fig_SQUID}, where the onset temperature is near 6.0 K.
The ZFC and FC measurements yield maximum signals of $\sim 95\%$
and $\sim 5 \%$ of complete flux expulsion, respectively.  The
values of $4 \pi \chi$ are accurate to roughly $\pm 15\%$ due to
uncertainties in mass, geometry and demagnetization factor of the
small crystal.  The small recovery of the diamagnetic signal on
field-cooling indicates either that the material shows strong vortex
pinning or that the superconductivity is not a
bulk phenomenon and is possibly associated with regions in the
crystal that are oxygen deficient. One such possibility is that
the superconductivity resides in a region near the surface which
has a composition different from that of the interior of the
crystal. The left inset to Fig.\ \ref{fig:fig_SQUID} shows $M$
versus $H$ for $T$ = 5 K, where the linear response of the
Meissner state up to $\sim 20$ Oe is followed by a decrease in the
magnitude of $M$ with increasing $H$, as is typical of flux
penetration in the vortex state of a type-II superconductor. The
$\chi(T)$ data in the normal state are shown in the right inset of
Fig.\ \ref{fig:fig_SQUID}. The susceptibility $\chi(T)$ increases
strongly with decreasing $T$ down to $\sim$ 220 K, where it
partially saturates. This temperature dependence is similar to
that seen for the compound Fe$_2$P \cite{wold_1970_1}, which may
be present as inclusions or surface impurities.  We estimate that
our observed magnetic susceptibility is consistent with 1-2\%
percent Fe$_2$P impurity.  It is notable that below $\sim$ 40 K,
$\chi(T)$ exhibits a weak upturn which persists down to 2 K. This
behavior is not typical for Fe$_2$P, and is either intrinsic to
LaFePO or due to a small concentration of some other paramagnetic
impurity.

\begin{figure}
    \begin{center}
        \includegraphics[width=12cm]{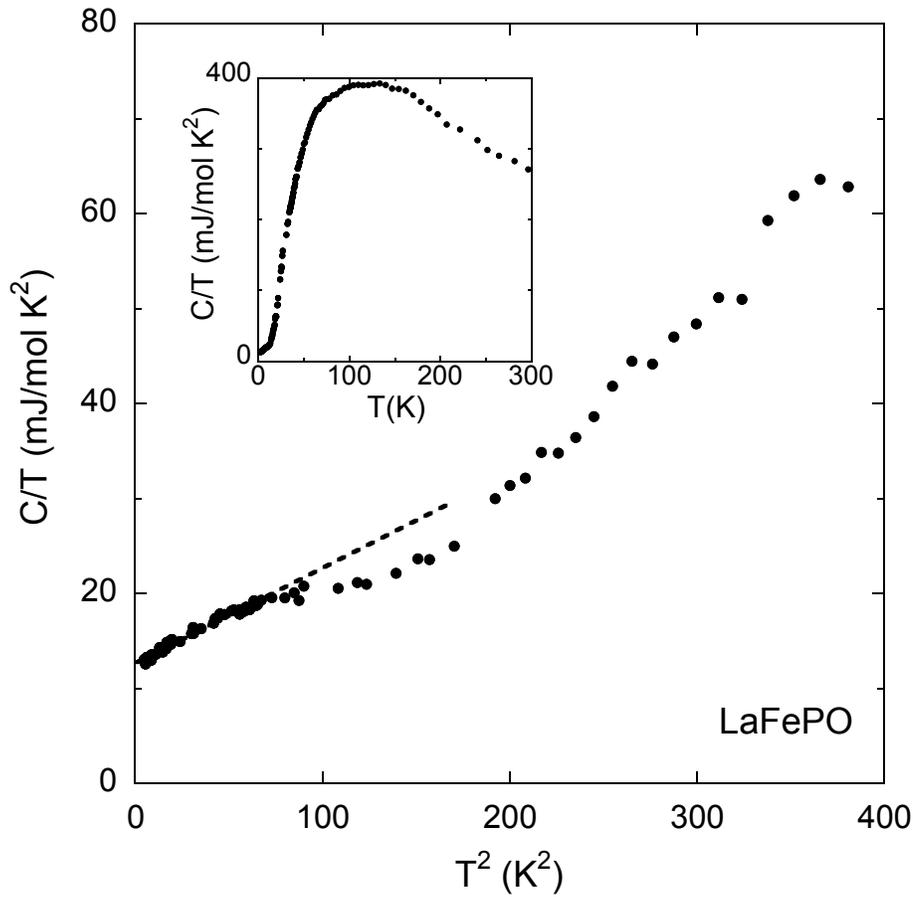}
    \end{center}
    \caption{Specific heat $C$ divided by temperature $T$, $C/T$, versus
    $T^2$ for 2-20 K. No discontinuity is observed near $T_c$, strongly suggesting that the superconductivity is not a bulk phenomenon. A linear fit (dashed line) to $C/T$ versus $T^2$ for 2-8 K yields an electronic specific heat coefficient $\gamma = 12.7$ mJ/mol$\cdot$K and a Debye temperature $\Theta _D = 268$ K. Inset: $C(T)/T$ versus $T$ from 2-300 K.}
    \label{fig:fig_Cp}
\end{figure}

Specific heat divided by temperature $C/T$ versus $T^2$ data for $2-20$ K are shown in Fig.\ \ref{fig:fig_Cp}. The absence of a detectable jump in $C(T)$ near $T_c \sim 6.6$ K, strongly suggests that the LaFePO single crystals do not exhibit bulk superconductivity. The specific heat jump at \Tc\ for a weak-coupling conventional superconductor is given by $\Delta C=1.52 \gamma T_c$ \cite{bardeen_1957_1}. For $T_c=6.6$ K and
$\gamma = 12.7$ mJ/mol$\cdot$K, this relation yields $\Delta
C=127$ mJ/mol$\cdot$K.  Comparing this expected specific heat
jump with the scatter in the $C(T)$ data (upper inset of Fig.
\ref{fig:fig_Cp}) would indicate that at most 10-20\% of the sample is
superconducting.  Our upper limit on the specific heat jump, $\Delta
C \lesssim 13$ mJ/mol$\cdot$K, is consistent with a very recent measurement of Kohama \textit{et al.\ } \cite{kohama_2008_1} on polycrystalline LaFePO, who find near $T_c$ a very small discontinuity of $\Delta C \sim 10$ mJ/mol$\cdot$K. These results support the interpretation that the
superconductivity is associated with defects such as oxygen
vacancies that act to dope the compound or possibly a different
surface composition in comparison with the bulk.  On the other hand,
LaFePO may exhibit some exotic type of superconductivity that is characterized by an unusually small energy gap.

The $C/T$ versus $T^2$ data shown in Fig.\ \ref{fig:fig_Cp} can be fit, between 2 K and 8 K, by the sum of electronic $\gamma T$ and lattice $\beta T^3$ terms, yielding an electronic specific heat coefficient $\gamma=12.7$ mJ/mol$\cdot$K$^2$ and a Debye temperature $\Theta_D=268$ K. The value of $\gamma$ is close to $\gamma=12.5$ mJ/mol$\cdot$K$^2$ obtained for a polycrystalline sample by McQueen \textit{et al.\ } \cite{mcqueen_2008_1}. From the value $\gamma=12.7$ mJ/mol$\cdot$K$^2$ and the saturated magnetic susceptibility $\chi_0$ $\sim$ 2 $\times$ 10$^{-3}$ emu/mol$\cdot$Oe, the Wilson - Sommerfeld ratio is calculated to be R$_W$ = 14.6. By comparison to the values expected for noninteracting electrons and electrons in a bound state Kondo singlet, R$_W= 1$ and 2, respectively, this value is unphysically large. This result indicates that the value of $\chi_0$ does not represent the Pauli paramagnetic susceptibility of the conduction electrons. For this reason, we consider the value of $\chi_0$ = 3 $\times$ 10$^{-4}$ emu/mol$\cdot$Oe reported by McQueen \textit{et al.\ }\cite{mcqueen_2008_1}, which yields R$_W$ = 2.19. This inconsistency is presumably due to the inclusion of 1-2\% Fe$_2$P, which would enhance the assumed low $T$ susceptibility for conduction electrons. From the coefficient $A = 9.46 \times 10^{-3} \mu \Omega$cm/K$^2$ obtained by fits to $\rho (T)$ and $\gamma = 12.7$ mJ/mol$\cdot$K$^2$, the ratio $R\equiv A/\gamma ^2 = 6\times 10^{-5}$ $\mu\Omega$ cm(mol$\cdot$K/mJ)$^2$ is calculated, in order of magnitude agreement with the Kadowaki - Woods value of $1\times 10^{-5}$ $\mu\Omega$ cm(mol$\cdot$K/mJ)$^2$ \cite{kadowaki_1986_1}.  As shown in the lower inset to Fig.\ \ref{fig:fig_Cp}, $C(T)/T$ continues to increase rapidly up to $\sim$ 120 K, where it goes through a peak and then decreases monotonically with increasing $T$.

\begin{figure}
    \begin{center}
        \includegraphics[width=12cm]{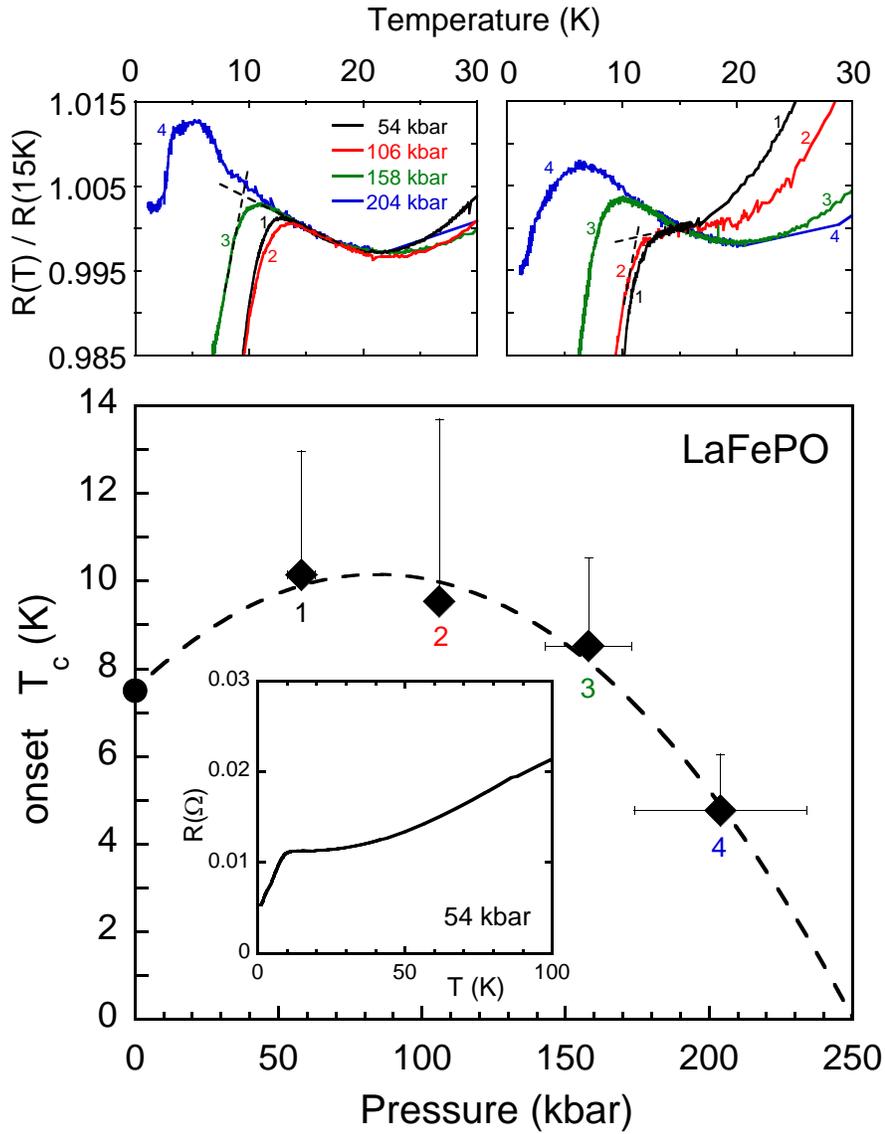}
    \end{center}
    \caption{(color) Top panels: Resistance curves normalized to the values at 15 K for two different lead configurations where the onset \Tc\ is defined as the intersection of the dashed lines. Main panel: Pressure dependence of the onset \Tc\ for P $=$ 54, 106, 158 and 204 kbar. The value of the $T_c$ onset at ambient pressure corresponds to the measurement of $R$ versus $T$ shown in Fig.\ \ref{fig:fig_R} with \Tc\ determined using the same criterion as that used for the high pressure points as described above. The high-pressure points correspond to the DAC measurements.  Horizontal bars represent the gradient of pressure along the sample, as described in the main text. The high temperature limit of the vertical bars represent an estimate of the first appearance of superconductivity, inferred from the first indication of the deviation of $\rho (T)$ from normal state behavior.  The dashed line is a guide to the eye.}
    \label{fig:fig_DAC}
\end{figure}

Electrical resistivity $\rho(T)$ measurements were made under high
pressures of 54, 106, 158, and 204 kbar, in that order, as shown
in Fig.\ \ref{fig:fig_DAC}. The upper panels in Fig.\ \ref{fig:fig_DAC} display the resistance versus pressure in the vicinity
of the superconducting transition for two different configurations
of the six tungsten leads.  Each different configuration measures
a somewhat different section of the sample.  The lower inset of
Fig.\ \ref{fig:fig_DAC} shows $\rho(T)$ over a broader temperature
range for the measurement at 54 kbar. The \Tc\ values are
determined graphically as illustrated by the dashed lines in the
upper panels and averaged over the two configurations of leads.
The values for the vertical bars represent the
superconducting onset temperature and are determined by the
temperature at which the resistivity passes through a maximum. The
pressure is determined from an average of the pressure given by
the two different pieces of ruby in the cell and the horizontal
bars represent the pressure difference determined from each
individual piece of ruby. The point at ambient pressure was taken
from the resistivity curve shown in Fig.\ \ref{fig:fig_R}.
Remarkably, a rather moderate pressure of 106 kbar is sufficient
to nearly double the onset of superconductivity from $\sim 7$ K to
$\sim 14$ K, above which pressure acts to suppress \Tc.

\section{Concluding Remarks}
Single crystals of the compound LaFePO prepared by a flux growth
technique at high temperatures exhibit a superconducting transition
near 6.6 K, as revealed by electrical resistivity and
magnetization measurements. The transition temperature \Tc\ is
suppressed anisotropically for magnetic fields perpendicular and
parallel to the ab plane, suggesting that the crystalline
anisotropy strongly influences the superconducting state.
Measurements of electrical resistivity under pressure show that
\Tc\ passes through a maximum of nearly 14 K at $\sim 110$ kbar,
indicating that significantly higher $T_c$ values may be possible
in the phosphorus - based oxypnictides. In contrast, specific
heat measurements show no indication of a discontinuity at \Tc,
suggesting that stoichiometric LaFePO does not exhibit bulk
superconductivity. These results indicate that either the
superconductivity is associated with oxygen vacancies that alter
the carrier concentration, or that the superconductivity is
characterized by an unusually small energy gap. Further experiments
in which the compound is doped with charge carriers through
chemical substitution will hopefully shed light on the
superconductivity displayed by this new class of superconducting
materials.
\textbf{Acknowledgements}
We thank S.\ T.\ Weir, D.\ D.\ Jackson, and Y.\ K.\ Vohra for
providing designer diamonds.  Thanks are also due J.\ R.\ Jeffries
for setting up our diamond anvil cell facilities and O.\ Shpyrko
for advice concerning x-ray diffraction.  Research at University
of California, San Diego, was supported by the U.S. Department of
Energy grant number DE-FG52-06NA26205, by the National Nuclear
Security Administration under the Stewardship Science Academic
Alliance Program through the U.S. Department of Energy grant
number DE-FG52-06NA26205, and the National Science Foundation
grant number DMR0802478.

\end{document}